\documentclass[preprint,aps,amsmath]{revtex4}
\usepackage{graphicx}

\begin{document} 

\title{Is a first order space-time theory possible?}

\author{Yaneer Bar-Yam}

\affiliation{New England Complex Systems Institute \\ 24 Mt. Auburn St., Cambridge, Massachusetts 02138}

\begin{abstract}
Einstein's general relativity relates the curvature of space time, a second order differential property, to the stress-energy-momentum tensor. In this paper we ask whether it is possible to develop a first order theory relating space-time angles to the energy-momentum vector. We suggest several concepts that would be relevant, including quantum mechanical concepts that are usually treated separately. Phenomenological Lorentz covariance arises from both field and coordinate transformation and the Dirac equation becomes a special case of the space-time field equation. We reinterpret Kaluza-Klein theory in this context by considering the compact fifth dimension as the quantum wavefunction phase. Further directions for development are suggested.
\end{abstract}

%\pacs{00.00}

\maketitle

General relativity [1,2] is conceptually based upon identifying gravitational acceleration as due to space curvature. Both acceleration and curvature are second order differential properties of space-time. Formally, the theory modifies the metric tensor of flat space-time, $g^{\mu \nu}$, which is defined in relation to the second order differential of space time distance $d^2s = g^{\mu \nu}dx_\mu dx_\nu$.  This second rank tensor is then related to the stress-energy-momentum tensor. We consider whether it might be possible to develop a first order theory that may be of interest in efforts to link general relativistic concepts with quantum fields. We discuss some ideas toward a first order theory to overcome some of the initial obstacles such as the need for a non-conventional origin of Lorentz covariance. At this point it is not certain that these steps are forced by the assumption of a first order theory, or whether there are other approaches that can be taken. 

A first order theory would start by assuming that the entity of interest is the velocity rather than the acceleration. The velocity would describe the relative angle of space-time axes rather than its curvature. This makes sense because if the time axis is at an angle with respect to the space axes of another coordinate system, giving a non-zero projection, over time the location of a reference spatial coordinate changes (Fig. 1). This corresponds to a translation of the coordinate systems relative to each other. Assuming the space and time axes are measured in the same units, the velocity vector $v$ consists of the negative of the directional cosins of the time axis of one coordinate system (the observer) with respect to another (the observed). 

A limitation on the magnitude of velocity $|v| \le 1$ follows, without additional assumptions, from the maximum projection of time onto space. While this is consistent with special relativity, we must nevertheless be concerned about the relevance of Lorentz transformations. The relationship between coordinate systems at an angle determined by the velocity does not correspond to the Lorentz transformation
\begin{equation}
\begin{array}{ll}
t &= \gamma (t_0  + v x_0) \\
x &= \gamma (x_0  + v t_0)
\end{array} 
\label{lorentz}
\end{equation}
where $ \gamma = 1/\sqrt{1-v^2}$. Instead it follows the equations (Fig. 1):
\begin{equation}
\begin{array}{ll}
t &= \gamma t_0 \\
x &= x_0 + v t = x_0 + \gamma v t_0 
\label{transform}
\end{array} 
\end{equation}
However, in a first order theory, we cannot consider the coordinate transformation separate from laws of mechanics describing matter. 
Conventionally we allow arbitrary observer velocities at any location in space-time. In a first order theory, velocity is associated with a state of matter at a location in space-time. An observer has a velocity because it is associated to a moving matter field, i.e. a moving observer is formed of matter. A way to think about a coordinate change is that in one region of space-time there is a matter field, the observer associated with that region extrapolates its local space-time axes to another region of space-time with a different local matter field. The extrapolation leads to a change of space-time angles which is related to the change of matter field between these locations. 
What is essential for electrodynamic phenomenology is that Lorentz covariance applies to physical law, i.e. as a property of the field equation, which includes both the change in the matter field and the change in coordinates. Thus, the physics should be described by a function of space-time,  a field $f(x_0,t_0;v),$ which when written in terms of the observer space-time coordinates $(x,t)$ is Lorentz covariant, so that the Lorentz relationship of velocity and coordinate change is consistent with the field expression. This can be guaranteed by insisting that the only combinations of coordinates that appear in the function $f$ are Lorentz covariant. For a Lorentz boost given by Eq. (\ref{lorentz})
a combination of the coordinates that is Lorentz covariant is readily identified to be 
\begin{equation}
\gamma (t-vx) = \gamma^2 (t_0 + v x_0 - v x_0 - v^2 t_0) = t_0 
\end{equation}
where in the original coordinate frame we take the velocity $v_0=0$ ($\gamma_0=1$) so that the expression on the right is equivalent to $\gamma_0 (t_0-v_0 x_0)$, which has the same form in the original coordinates as the expression on the left in the new coordinates, implying Lorentz covariance. Inserting the coordinate transformation of Eq. (\ref{transform}) we have that
\begin{equation}
%\begin{array}{ll}
\gamma (t - vx) = \gamma (\gamma t_0 - v x_0 - v^2 \gamma t_0) = t_0  -  \gamma v  x_0
\label{tvx}
%\end{array} 
\end{equation}
This is therefore the dependence of the argument of $f(x_0,t_0;v)$ or $f(x,t;v)$ that is needed to give Lorentz covariance for the field. Still, this is not the formally correct expression since it itself depends on $v$ whereas it should depend upon a property of matter. In this context, $v$ is defined by the angle of space and time, a property relating two coordinate systems, not a property of matter. We must substitute the variable describing the matter field that is related to $v$. We will obtain the necessary expression below. 

\begin{figure}
\includegraphics[width=12cm]{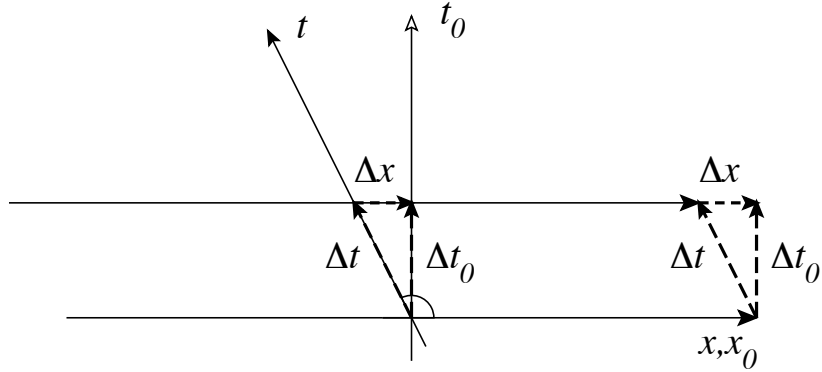}
\label{fig1}
\caption{Illustration of an example of an observer coordinate system $(t,x)$ for which the direction of time is not orthogonal to space. Two horizontal (space) axes are shown differing by observer time interval $\Delta t$ and local time $\Delta t_0$. The angle of time can be described by the negative of the directional cosins that are the velocity of the local orthogonal coordinate axes $(t_0,x_0)$ with respect to the observer. Explicitly, consider the angle indicated by the arc between the diagonal time direction, $t$, and the horizontal space axis $x,x_0$. The cosign of this angle is $-\Delta x / \Delta t =- v$, and this gives one of the coordinate transformation equations as $x=x_0+v t$. We also have  $\Delta t_0   = \sqrt{ \Delta t^2-\Delta x^2}=\gamma^{-1}\Delta t$, giving the other coordinate transformation equation $ t   = \gamma t_0$.  }
\end{figure}

The next step is to identify a property of matter associated with a location in space that determines the coordinate system angle. To reduce the order from that of the second rank stress-energy-momentum tensor, which in the simplest case is the mass density, we might consider taking a ``square root" of the density. While this is not easy to understand from a classical perspective, in a quantum formalism a natural assumption would consider the wavefunction as the origin of space-time bending, since the wavefunction amplitude squared is the density. Indeed, in the absence of any other physical variable, it seems reasonable to propose that the quantum wavefunction itself be considered as a description of the space-time structure. To have a local vector property that gives rise to the velocity, we consider the momentum operator on the wavefunction. Thus we assume that the momentum is the property of matter that indicates the space-time angle that is revealed by the velocity. Classically, the relationship between momentum and velocity is simple, here this relationship would be reinterpreted as a relationship of a property of matter with a property of space-time geometry---conceptually analogous to the Einstein field equations that relate properties of mater to space-time curvature. Rewriting Eq. (\ref{tvx}) in terms of the relativistic expressions for momentum, $k=m\gamma v$, and energy, $\omega = m \gamma$, it is natural to multiply by particle rest mass $m$ (a constant factor does not affect the Lorentz covariance) we have 
\begin{equation}
\omega (t - vx) =  \omega t - k x  =  m t_0  -  k  x_0 .
\end{equation}
We recognize the second expression as the space-time variation of the phase of a conventional plane wave. Our analysis indicates that we must use $z=m t_0 - k x_0$ to describe the variation of a field function in space so as to obtain phenomenological Lorentz covariance. This can be interpreted by saying that the only variable that the field depends on varies at a fixed rate, $\partial_{t_0} z = m$, with the direction perpendicular to space, and varies in space at a rate given by the momentum, $-\partial_{x_0} z = k$. The velocity, and therefore the direction of time, is then given by the usual relativistic relationship between the momentum and velocity $v=k/\omega$, where $\omega$, defined as $\sqrt{k^2+m^2}$, is the rate of change of the field variable with time, $d_{t} z =\omega$, i.e. the path derivative in the direction of time. 

From the expression for $\omega$ it is possible to see that the direction of time is the direction in Euclidean $(x_0,t_0)$ space that has the maximum rate of change of the field variable. This is appealing as an extremum principle. If we treat $k$ and $v$ as independent variables (as justified by allowing the direction of time to vary, with $v$ the negative of the directional cosins of the direction of time, and $k$ the negative of the field gradient in space) and we define the generalized Hamiltonian function as the rate of change of the field variable in the direction of time for arbitrary directions of time,
\begin{equation}
H(k,v) =   k v + m \sqrt{1+v^2}  ,
\end{equation}
then the maximum of $H(k,v)$ with respect to variations of $v$ keeping $k$ constant gives the Hamiltonian $H(k)=\omega$. If we allow $k$ to be a function of $x$ (a matter field with a spatially varying momentum) it is possible to show that this maximization leads to Hamilton's and Lagrange's equations of motion, and is therefore equivalent to (relativistic versions of) traditional extremum principles for classical mechanics. Explicitly, a variable $k$ implies  $dH(k)/dk = \partial H(k,v)/\partial k + (\partial H(k,v)/\partial v)(dv/dk) = \partial H(k,v)/\partial k = v$ (the second equality arises from the variational time direction determination as given by $\partial H(k,v)/\partial v = 0$, and the final equality from the explicit form of $H(k,v)$). This is the first of Hamilton's equations. Since $k$ varies with $x$, then $v(k)$, and $H(k)$, are functions of $x$. The time derivative of $k$ is given by $dk/dt = (dk(x)/dx)(dx/dt) = (dk/dx)(dH(k)/dk) = dH(k)/dx$. This is a positional change in the observed `kinetic energy.' Any change of $k$ must be due to an extrinsic cause, `The Force.' Specifically, we assume a conserved `total energy' $H(k,x) = H(k) + \phi (x)$, which is the combined rate of state change of the system and environment (or system and other system) implying $d\phi /dx =  - d\omega / dx$, and $dk / dt =  - d\phi / dx =  - \partial H / \partial x$, which is the second of Hamilton's equations. For the Lagrangian we define $L(k,v) =  k v - H(k,v)$. The explicit form of $H(k,v)$ implies that $L(k,v)$ does not depend on $k$ directly, and therefore can be written as $L(v)$, but the determination of the direction in time is equivalent to $dL(v)/dv = k$. Lagrange equations of motion follow from $L(v,x) = L(v) - \phi (x)$ as $ (d/dt) \partial L(v,x)/\partial v = \partial L(v,x)/\partial x$. Thus Hamilton's and Lagrange's equations can be inferred from the determination of the direction of time as a maximum of the rate of change of field variable, where the non-orthogonality of space and time plays a role in the transformations. 

Any function of $z$ as defined above can be used to form a field variable and will be Lorentz covariant. We will explore such a field theory where the field is cyclical and can be represented as $\psi=e^{-iz}=e^{i(kx_0-m t_0)}=e^{i(kx-\omega t)}$.\cite{ZM1,ZM2,ZM3,ZM4}  
This allows a direct connection with quantum mechanics and the Dirac equation.

To define a first order field equation in a traditional form, a differential equation relating the time dependence (dynamics) to other system properties would be appropriate. Since the time direction is at an angle to the space directions, observer time can be treated as a dependent variable, which depends on space, and the direction orthogonal to space, $t_0$. Specifically, the time derivative can be related to a function (the Hamiltonian in a new form) of space derivatives (the momentum) and $t_0$. Thus, we are looking for an equation that relates time and space derivatives of the form
\begin{equation}
 d_t   =  v \partial _{x}   +  \sqrt{1+v^2} \partial _{t_0}   
\label{fieldeq0}
\end{equation}
Since this is an intrinsic relationship of derivatives, a field variable that describes matter should satisfy this equation.
A candidate first order field equation is the Dirac equation. [3-5]
\begin{equation}
 id_t \Psi  = -i\alpha \partial _{x} \Psi  + \beta m  \Psi  
\end{equation}
Where $\alpha$ and $\beta$ are Dirac matrices, and $\Psi$ is a Dirac wavefunction. To make the correspondence more direct, we consider the mass term of the Dirac equation to be a derivative of the wavefunction with respect to $t_0$, i.e. the local frame of reference, $m \Psi  = i \partial_{t_0} \Psi$. 
The Dirac equation then appears to be a transformation equation of the space-time derivatives between reference frames. The time derivative in the observer frame of reference is related to the space and $t_0$ derivatives, where the $t_0$ `rest frame' derivative gives the rest mass of the particle. There are many issues to address. We start by addressing the existence of Dirac matrices and spinors and the role of spin and anti-particle degrees of freedom. 

The Dirac spinor has well known properties as a representation of Lorentz transformations. Here we consider this as capturing the coordinate transformation between two locations in space. Boosts are mildly reinterpreted as giving the angle of the time coordinate, otherwise the interpretation is similar to the conventional treatment of coordinate transformations. Specifically, Dirac equation eigenfunctions in their traditional forms [3-5]
can be represented, and are here interpreted as a product $\Psi =  \chi \psi$. $\chi$ is a constant Dirac spinor which is a representation of a coordinate transformation consisting of the angle of time given by $v$ and a rigid rotation, and possible time and space coordinate inversions. $\psi$ is a momentum and energy eigenstate plane wave which represents the field variable as indicated previously. Considering only the eigenvectors, we multiply the Dirac equation on the left by $\chi^\dagger$ to obtain 
\begin{equation}
 i\chi^\dagger d_t \chi \psi = -i\sum_i \chi^\dagger \alpha _i \partial _{x_i } \chi \psi  + \chi^\dagger \beta m  \chi \psi  
\end{equation}
since $\chi$ is constant in space-time we obtain (after dividing by $\chi^\dagger\chi$)
\begin{equation}
 i d_t \psi = -i v  \partial _{x }  \psi  + \gamma^{-1}  m \psi  
\end{equation}
Considering the right term as given by the partial derivative with respect to $t_0$ we obtain the desired Eq. (\ref{fieldeq0}). 

The appearance of antiparticles and spin in Dirac's equation is conventionally remarkable because the framing of the Dirac equation is as a description of a structureless point particle. In our case, however, the Dirac equation is a field equation relating how an observer in one region of space (region $A$) extrapolates a definition of time and space to another region of space (region $B$). The relationship of the two space-time coordinates (native and extrapolated) in region $B$ can then be identified through the behavior of the field equation.  This relationship between two coordinate frames has more structure than a point object and gives reason for the existence of particle, antiparticle, and spin states. Specifically, if we assume the possibility of axis inversions in the relationship between coordinates associated with different regions of space, there are four distinct sets of states. These are related by inversion of the direction of time and inversion of a spatial axis, which are distinct and neither of which can be achieved by continuous transformations. Considering the inversion of time as giving rise to antiparticles is consistent with Feynman's [6] interpretation of an antiparticle as moving backwards in time, without the filling of a negative energy sea. We note that counter to conventional expectations, inverting the time axis does not change the sign of the momentum $k$---which is the negative gradient of the field variable in space and thus does not depend on the time axis---but does change the sign of $\omega$, and the sign of $v=\omega/k$, so that $v$ and $k$ are in opposite directions for antiparticles. In traditional discussions of the Dirac equation, where the same issue arises, it is standard to invert the definition of momentum for antiparticles to be the gradient (rather than the negative gradient) of the wavefunction. This change of definition leads to a more intuitive situation where the velocity is in the same direction as the momentum. Considering the inversion of spatial axis as giving rise to spin is consistent with the transformation properties of the Dirac equation eigenvectors. Coordinate inversion results in parity inversion of the relationship of the axes as is found for spin inversion. Moreover, for the standard Dirac equation eigenvectors, it can be verified that axis inversions perpendicular to the momentum (that do not change the velocity or momentum) change the spin coordinate. Axis inversions that change the momentum and velocity must include a change of the labeling of states and therefore the relationship of spin states is not as transparent.

The treatment thus far considers only Dirac equation eigenfunctions that are plane waves. Wave function superposition may be reinterpreted as a case of a multiple sheeted space time, i.e. one where extrapolation from one location in space to another results in multiple passes over the same region of space due to severe bending. We defer to later work most issues dealing with multiple particles including interaction fields, Maxwell's equations, particle statistics and measurement theory. We only note in this context  the introduction of minimal coupling of the vector potential through definition of the canonical momentum in the Dirac equation. The argument is essentially similar to the concept of gauge invariance as it postulates that the observer can redefine the field variable (wavefunction phase) by an analytic function of space-time and still describe the same system. Specifically, the Dirac equation without a field corresponds to a fixed direction of time over the region of space-time to which it applies. Variations of the time direction in space-time would result from the coupling of multiple field components, when separating a particular field component and treating it as a separate field. The effect of the excluded components would then be described at some level of approximation by the electromagnetic field. The direction of time of the observer, as determined by the entire system, is different from the direction of time indicated by the gradient of the component field. The same issue arises when we consider a system formed of only a single component if we allow the observer to vary its definition of the direction of time over space, in effect separating some of the behavior of the system to the environment. Thus, the introduction of the electromagnetic potential $A$ can be seen through the need to allow an observer to adopt a different direction of time than that associated to a field. In this case the field variation $k=-\partial_x z$ would not be equivalent to the field variation that would be needed to give rise to the time direction specified by the observer. We define the field variation that would be needed to give the observer time direction as the canonical momentum $\pi$. The difference between the canonical and field component momentum is the variation ascribed to external influence $A=k-\pi$. Similarly, the variation of the field in time that would be expected from the field equation $\pi_0$ is not that which would be obtained from the field component variation in time $\omega = \partial_t z$ and the difference $\phi=\omega-\pi_0$ is the scalar  potential. By choosing a convention of the direction of time over space-time, the observer is partitioning the variation of the field variable $z$ that is part of the system and that that is not. 
A gauge transformation corresponds to the observer changing the reference direction of time over space (as well as coordinate rotations), where limitations are placed on the arbitrariness of the variation over space in assuming phase changes that are analytic single valued functions of space-time.

As a last step toward a first order theory taken in this paper we use these concepts to reinterpret Kaluza Klein (KK) theory.[7-11] Here, we restrict our discussion to the original concepts of KK theory and not the many extensions. KK theory follows the framework of general relativity in postulating a metric for space-time, but considers a five dimensional space time with four space-like dimensions.  A particular postulated metric in five dimensions without sources can be used to obtain general relativity with a source term in four dimensional space-time. The absence of observation of a fifth dimension corresponding to a fourth spatial dimension is most commonly treated by considering it to be circularly compact and small in size compared to scales of observation. 

We reinterpret KK theory by assuming that the fifth dimension is the field variable $z/m$ which leads to the following changes of perspective. First, $z/m$ can be considered as not part of the observer's space-time dimensions and therefore does not impact on the dimensionality of space as seen by the observer. Second, where $z/m$ does appear in the formalism it can be considered to be circular and compact with a radius of $1/m$, because it appears only in the wave function as $e^{-iz}$. Third, at this point the reference mass, and thus the length scale could be for any particle, from an electron, well within the range of physical observation,  to the plank mass. Fourth, $z/m$ is in a sense both observable, through well-known quantum effects, and non-observable, because it is the phase of a quantum mechanical wave function. Fifth, the intrinsic relationships between $z/m$ and the space-time dimensions results in a dependency between the five dimensions $(t,x,z/m)$, that can be understood as a metric of the effective five dimensional space-time. In the next paragraph we show that this metric is consistent with KK theory. In summary, in this context, there is no problem with the existence of such a dimension since where it is relevant to be observed it is observed, and it does not have the same properties as a normal space dimension.

Interestingly, the relationship we have of the angle of time in the five dimensional space $(t,x,z/m)$ can give rise to the metric of KK theory. We calculate the metric starting from the expression $ds^2 = dx_i  dx^i = dt^2  - dx^2  - d(z/m)^2 = g_{i j}dx^idx^j$, using summation convention for repeated indices, where i,j runs from 0 to 4, with 0 to 3 corresponding to the usual four dimensional space time captured by the indices $\mu,\nu$. The first two terms $dt^2  - dx^2$  yield the usual metric $g_{\mu \nu}$. The term $d(z/m)^2$ can be considered to make contributions to space-time, as well as to the metric of the fifth dimension. From the field variable properties related to the angle of time we have $dz  = \pi_0 dt - \pi dx = \pi_\mu dx^\mu$. Thus we can write the contribution of the fifth dimension to the metric in three ways, $-d(z/m)^2 = -d(z/m)(\pi_\mu/m) dx^\mu=-(\pi_\mu /m) (\pi_\nu /m) dx^\mu dx^\nu$. We choose to include all three by adding the first and third and subtracting the second to obtain the metric:
\begin{equation}
g = \left( {\begin{array}{*{20}c}  {g_{\mu \nu }  - (\pi _\mu  /m)(\pi _\nu  /m)} & { (\pi _\mu  /m)}  \\  { (\pi _\nu  /m)} & -1  \\
 \end{array} } \right)
 \label{metric}
\end{equation}
which corresponds in form to the KK metric, though the canonical momentum appears instead of the vector potential. Considering the momentum in general requires treating multiple component (particle) aggregation. However, by gauge invariance, a uniform momentum can be eliminated from the space-time metric. Thus, under circumstances where we can consider the field equation to describe a uniform test particle, the canonical momentum can be replaced by the vector potential, as is found in the traditional form of the KK metric. This is consistent with the recognition that fixed angles of space-time do not affect the curvature of the metric. 
We note that the metric of the fifth dimension has unit magnitude as was assumed in the original KK theory. Further, we note that conventionally the metric signature of the fifth dimension is considered as being space-like. In our case $t_0$ has a space-like signature even though it is related to a time coordinate. This is because the `time-like' signature of $t$ arises as a result of being the observer time rather than as a characteristic  of time coordinates.

In this paper we have shown that the question of whether it might be possible to develop a first order space-time theory suggests interesting reinterpretations of existing physical laws. The concept of combining quantum mechanics with the general relativistic formalism in this way invites considering many additional issues. Whether it is possible to develop a first order space-time theory depends on providing answers to a variety of open questions related to general relativity and quantum theory that are not yet addressed here. Further, to confirm that such a theory is useful it must provide predictions that are not the same as conventional theories. This is a challenge that remains to be met.

In summary, in Einstein's general relativity acceleration is associated with a location through the curvature of space-time. Here we considered whether it might be possible to formulate a theory in which velocity is the entity of interest. Velocity is associated with a particular location through the momentum of the quantum wavefunction. Velocity specifies the angle of time with respect to space. Further work is needed to show that a complete first order theory can be developed.

\end{document}